\documentstyle[preprint,aps]{revtex}
\begin{document}
\baselineskip15truept
\draft

\title{Discontinuous Interface Depinning from a Rough Wall}

\author{G. Giugliarelli}
\address{Dipartimento di Fisica e INFN (Gruppo Collegato di Udine,
Sezione di Trieste), Universit\'a di Udine, 33100 Udine, Italy}

\author{A. L. Stella}
\address{INFM -- Dipartimento di Fisica e Sezione INFN, Universit\'a
di Padova, 35131 Padova, Italy}

\maketitle

\begin{abstract}
Depinning of an interface from a random self--affine substrate with
roughness exponent $\zeta_S$ is studied in systems with short--range
interactions. In 2$D$ transfer matrix results show that for
$\zeta_S<1/2$ depinning falls in the universality class of the flat
case. When $\zeta_S$ exceeds the roughness ($\zeta_0=1/2$) of the
interface in the bulk, geometrical disorder becomes relevant and,
moreover, depinning becomes \underline{discontinuous}. The same
unexpected scenario, and a precise location of the associated
tricritical point, are obtained for a simplified hierarchical model.
It is inferred that, in 3$D$, with $\zeta_0=0$, depinning turns
first--order already for $\zeta_S>0$. Thus critical wetting may be
impossible to observe on rough substrates.
\end{abstract}


Wetting and depinning phenomena occur when the interface between two
coexisting phases unbinds from an attractive substrate
\cite{forgacs}. In 2$D$, where an Ising interface can be well
represented by a directed path \cite{chui}, critical wetting occurs
as a rule, and first--order wetting is predicted only for special
setups \cite{forgacs}. In 3$D$, on the other hand, numerical
simulations have shown critical, first--order and tricritical wetting
for the Ising model. The same phenomena in random media have attracted
a lot of attention recently. Many works
\cite{huse,kardar&kardar,lipowski} considered the effects of quenched
disorder due to \underline{bulk} impurities on interface depinning.
Other studies addressed disorder restricted to a geometrically smooth
\underline{surface} (chemical surface disorder) \cite{forgacs1}. Both
bulk and chemical surface disorder may modify the universality class
of the unbinding transition, but generally not its continuous,
second--order character in 2$D$.\cite{forgacs}

In the present Letter we address the role of \underline{geometrical}
disorder due to wall roughness in determining the nature of the
wetting transition. Disordered geometry is most amenable to
theoretical treatment when characterized by simple scaling laws, like
in the case of self--affine or fractal substrates. So far, the
effects of such roughness were seldom discussed, mostly in connection
with complete wetting \cite{li&kardar,pfeifer,giugliarelli}. In spite
of this, real substrates with self--affine geometry are met in many
situations and have been the object of recent experiments
\cite{pfeifer}. For these substrates the average transverse width of
the wall scales with the longitudinal length $X$ as $X^{\zeta_S}$
($0<\zeta_S<1$).

We show that in the 2$D$ Ising model the nature of the depinning
transition changes drastically upon increasing the self--affine
roughness exponent $\zeta_S$. For $\zeta_S\lesssim1/2$ depinning
remains continuous as in the flat case, and disorder is irrelevant.
For $\zeta_S\gtrsim 1/2$ an unusual \underline{discontinuous}
depinning occurs. Since the intrinsic roughness of a 2$D$ interface
is $\zeta_0=1/2$, it is natural to expect $\zeta_S=1/2$ as precise
threshold. The emerging scenario is that of geometrical disorder
determining a tricritical phenomenon for depinning. We are able to
locate the tricritical point by accurate renormalization group (RG)
calculations on a hierarchical model, which further suggest
$\zeta_S=1/2$ as the tricritical threshold in 2$D$. Extrapolation of
our results to the experimentally most relevant case of 3$D$, where
$\zeta_0=0$ in ordered bulk, suggests that a minute disorder in
substrate geometry could suppress critical wetting in favor of a
first--order transition.

\vskip6.0truecm
\includegraphics{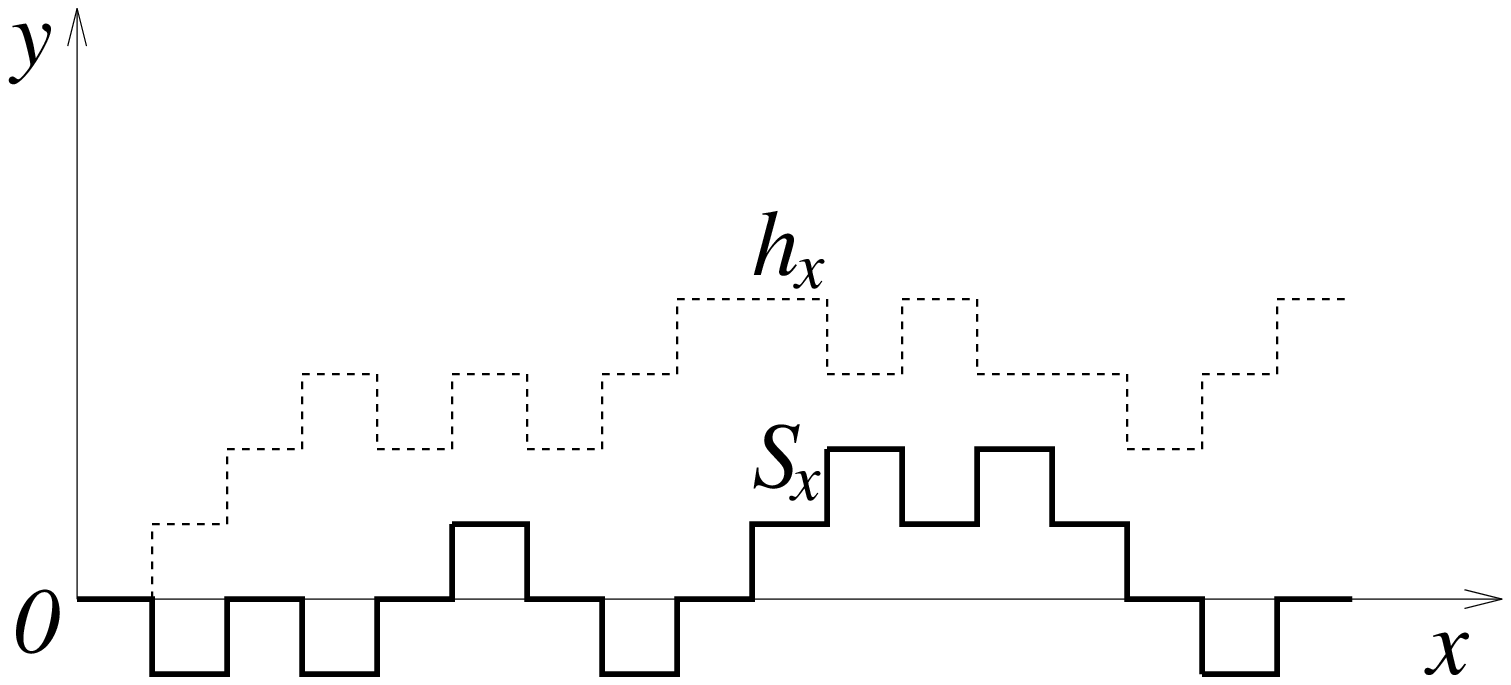}
\noindent{\bf FIG.1.} Example of rough substrate boundary (continuous)
and interface (dotted) configuration.
\bigskip

   Let $x$ and $y$ be integer coordinates of points on a square
lattice. A ``{\sl wall}'' is given by a self--avoiding path directed
according to the positive $x$ axis (Fig. 1). To each wall step
parallel to the $x$ axis, a height $S_x$, equal to the ordinate of
its left end, $x$, is associated. For simplicity we consider only
wall configurations obeying the restriction: $s_x=S_x-S_{x+1}=\pm 1$.
In order to generate wall geometries with a preassigned roughness
exponent $\zeta_S$, we used a randomized version of an algorithm due
to Mandelbrot \cite{mandelbrot}. Given a wall profile, the interface
can assume configurations obeying $h_x-h_{x+1}=0,\pm 1$ and $h_x\geq
S_x$, if $h_x$ is the height of the horizontal step at $x$ (Fig. 1).
The interface is like a (partially) directed walk or polymer, and its
Hamiltonian is
\begin{equation}
{\cal H}=\sum_x [\varepsilon(1+|z_x-z_{x+1}+s_x|)-u\delta_{z_x,0}]
\eqnum{1}
\end{equation}
where $z_x\equiv h_x-S_x$, and $\varepsilon, u>0$. According to (1),
at temperature $T$, fugacities $\omega\equiv e^{-\varepsilon/T}$ and
$k\equiv e^{u/T}$ are associated to each (horizontal or vertical)
step of the walk, and to each horizontal step on the wall,
respectively. With a wall profile covering a distance $X$ the
partition function is
\begin{equation}
{\cal Z}_X=\sum_{all\ walks} \omega^L k^l
\eqnum{2}
\end{equation}
where the sum is restricted, e.g., to walks from the origin $(0,0)$
to any point $(X,y)$, with $y\geq S_X$. $L$ and $l$ indicate total
length and number of horizontal steps on the wall, respectively.
${\cal Z}_X$ is a functional of the wall profile. In order to
calculate it we use transfer matrices:
\begin{equation}
({\bf T}_{s_x})_{m,n}=\omega
[\delta_{n,m\!-\!s_x}\!+\!(\delta_{n,m\!-\!s_x\!-\!1}\!+\!
	      \delta_{n,m\!-s_x\!+\!1})\omega] k^{\delta_{n,0}}
\eqnum{3}
\end{equation}
where $m$ and $n$ range on the allowed $z_x$ and $z_{x+1}$,
respectively. The partition function thus becomes
\begin{equation}
{\cal Z}_X=\sum_{l,p} \left( \prod_{x=0}^{X-1} {\bf T}_{s_x}\right
)_{l,p} \phi_0(p)
\eqnum{4}
\end{equation}
where, with the left end of the interface grafted at the origin,
$\phi_0(p)=\delta_{p,0}$. A wall profile corresponds to a sequence of
factors ${\bf T}_1$, ${\bf T}_{-1}$ in the product of eq.(4). ${\cal
Z}_X$ is thus given asymptotically in terms of the largest Lyapunov
eigenvalue \cite{crisanti}
\begin{equation}
\lambda_{max}=\!\lim_{X\to\infty}\!\left
[\!{{||\left(\prod_{x=0}^{X-1} {\bf T}_{s_x}\right) {\vec \phi}_0||}
\over {|| {\vec \phi}_0||}}\! \right ]^{1\over X}\!\!=\! \lim_{X\to\infty}
\!\left[\!{{||{\vec \phi}_X||}\over {||{\vec \phi}_0||}}\!\right] ^{1\over X}
\eqnum{5}
\end{equation}
We verified that different long ${\bf T}$ sequences, i.e. wall
profiles, lead to the same eigenvalue within good accuracy. Thus, the
quenched free energy is $\ln \lambda_{max}(\omega,k,\zeta_S)=
\lim_{X\to\infty} \overline{\ln {\cal Z}_X}/X$, where the bar
indicates quenched averaging over wall profiles. In the random
context depinning is most efficiently detected by studying the
behavior of quantities which can be directly related to the
components of ${\vec{\phi}}_X$. Examples are the average probability,
$\overline{P_0}(\omega,k,\zeta_S)=\lim_{X\to\infty}\overline{{1\over
X} \sum_{x=0}^{X-1} \phi_x^2(0)/||\vec{\phi}_x||^2}$, that the
horizontal step lies on the wall, and the average distance of the
interface from the substrate, $\overline{\langle
z\rangle}=\lim_{X\to\infty}\overline{ {1\over X} \sum_{x=0}^{X-1}
\sum_z z\phi_x^2(z)/||\vec{\phi}_x||^2}$ \cite{forgacs}. For our
determinations we used up to 50 independent profiles with $x\leq
2^{20}$ for which the components of $\vec{\phi}_x$ were computed up to
a distance from the wall $z_{max}\simeq 2\cdot 10^4$.
\newpage

\noindent TABLE I. Values of $\psi$ and $k_c$ at $\omega=1/2$
from fits of $\overline{\langle z\rangle}$.
\begin{quasitable}
\begin{tabular}{ccc}
\hline
$\zeta_S$ & $\psi$ & $k_c$ \\
\tableline
0 & 1 & 4/3$^{\ a}$ \\
1/3 & 1.01 $\pm$ 0.01 & 1.772 $\pm$ 0.001 \\
2/5 & 1.01 $\pm$ 0.02 & 1.828 $\pm$ 0.001 \\
1/2 & 1.03 $\pm$ 0.04 & 1.908 $\pm$ 0.003 \\
\hline
\end{tabular}
\end{quasitable}
$^{\ a}\ $ Exact results for a flat substrate [1].
\bigskip

   Rather than considering variations of  $\overline{P_0}$ or
$\overline{\langle z\rangle}$ along curves parametrized by
temperature, we choose to follow $\omega=const$. lines. In the case
$\zeta_S\lesssim 1/2$, e.g., upon approaching $k=k_c$ from above with
$0<\omega<1$, $\overline{\langle z\rangle}$  is well fitted by
$\overline{\langle z\rangle}=A(k-k_c)^{-\psi}+B$, with $\psi$ always
compatible with the exactly known flat wall value of 1
\cite{forgacs}.  $k_c$ of course depends on $\omega$ and $\zeta_S$.
Some $\psi$ and $k_c$ determinations are reported in Table I for
$\omega=1/2$. For $\zeta_S\lesssim 1/2$ disorder in the wall geometry
does not appear to lead to a new universality class for depinning. By
treating a different model with continuum many--body techniques, Li
and Kardar \cite{li&kardar} found second--order depinning with
$\psi=1$ for all $\zeta_S<1$.  At variance with this conclusion we
find here that the situation drastically changes for $\zeta_S\gtrsim
1/2$. In this range $\overline{\langle z\rangle}$ has a much steeper,
abrupt rise at $k_c$, so that the previous fit becomes clearly
inappropriate.

  Further strong evidence of a change of the nature of the transition
at $\zeta_S\simeq 1/2$ comes from the behavior of $\overline{P_0}$
(Fig. 2). While for $\zeta_S\lesssim1/2$, $\overline{P_0}\sim
(k-k_c)^\rho$, for $k\to k_c^+$, with $\rho\simeq 1$, as with flat
substrate \cite{forgacs}, for $\zeta_S\gtrsim 1/2$ a discontinuity in
$\overline{P_0}$ shows up (Fig. 2), becoming sharper and sharper as
finite--size effects are reduced. The amplitude of the discontinuous
jump in $\overline{P_0}$ also increases with $\zeta_S$.

\vskip8.5truecm
\includegraphics{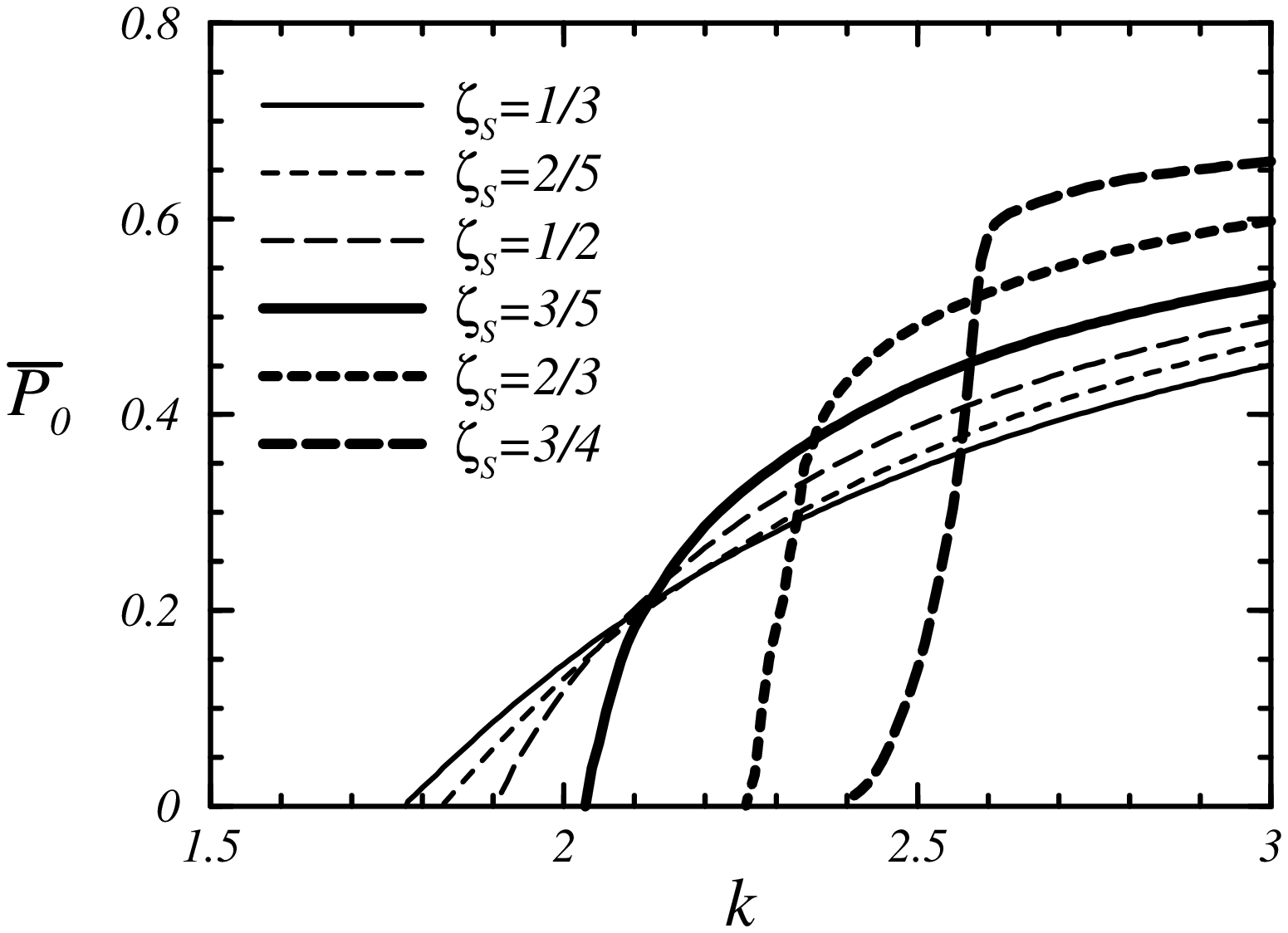}
\noindent{FIG. 2.} The probability $\overline{P_0}$ as a
function of $k$ at $\omega=1/2$ and for different $\zeta_S$ values.
\bigskip

Thus, for $\zeta_S\gtrsim 1/2$ substrate roughness is relevant and,
moreover, drives the depinning transition first--order. $\zeta_S=1/2$
is the natural candidate as border value between continuous and
discontinuous regimes. Indeed, for $\zeta_S>1/2$ the wall roughness
exceeds the roughness $\zeta_0=1/2$ \cite{lipowski} of the interface.

In 2$D$, first--order depinning is quite unexpected in the context of
interfacial phenomena. Only two special ways of obtaining it have
been conceived so far, by introducing either an attractive defect
line in the bulk \cite{forgacs2}, or longitudinally fully correlated
disorder \cite{nieu}. Here first order is caused by sufficiently
strong geometrical surface disorder, which also reveals opposite in
its effects to its chemical counterpart. Indeed, while higher
roughness induces first--order, in the defect line case chemical
surface disorder drives the transition back continuous
\cite{forgacs3}. This latter effect is certainly what one would
expect at first sight on the basis of experience with phase
transitions \cite{hui}. Like in the special examples of refs.
\cite{forgacs2,nieu}, our first order depinning needs not be
accompanied by off-coexistence prewetting phenomena.

Extrapolation to 3$D$ of our findings is natural and has remarkable
and unexpected consequences. Systems with a dominance of short--range
interactions are, e.g., metallic substrate--adsorbates or, even more,
type--I superconductors \cite{okki}. Since typically interfaces in
pure systems are only logarithmically rough in 3$D$ ($\zeta_0=0$), a
minimum of substrate roughness should be sufficient to give
first--order wetting. This offers a further possible explanation of
the fact that critical wetting is so elusive from the experimental
point of view. A most recent work predicts critical wetting for
superconductor interfaces \cite{okki}. For such interfaces $\zeta_0$
is not known, unfortunately, but could be rather small, if not zero.
This means that special care in using smooth substrates should be
exerced, in order to observe the predicted phenomenon.

\vskip6.0truecm
\includegraphics{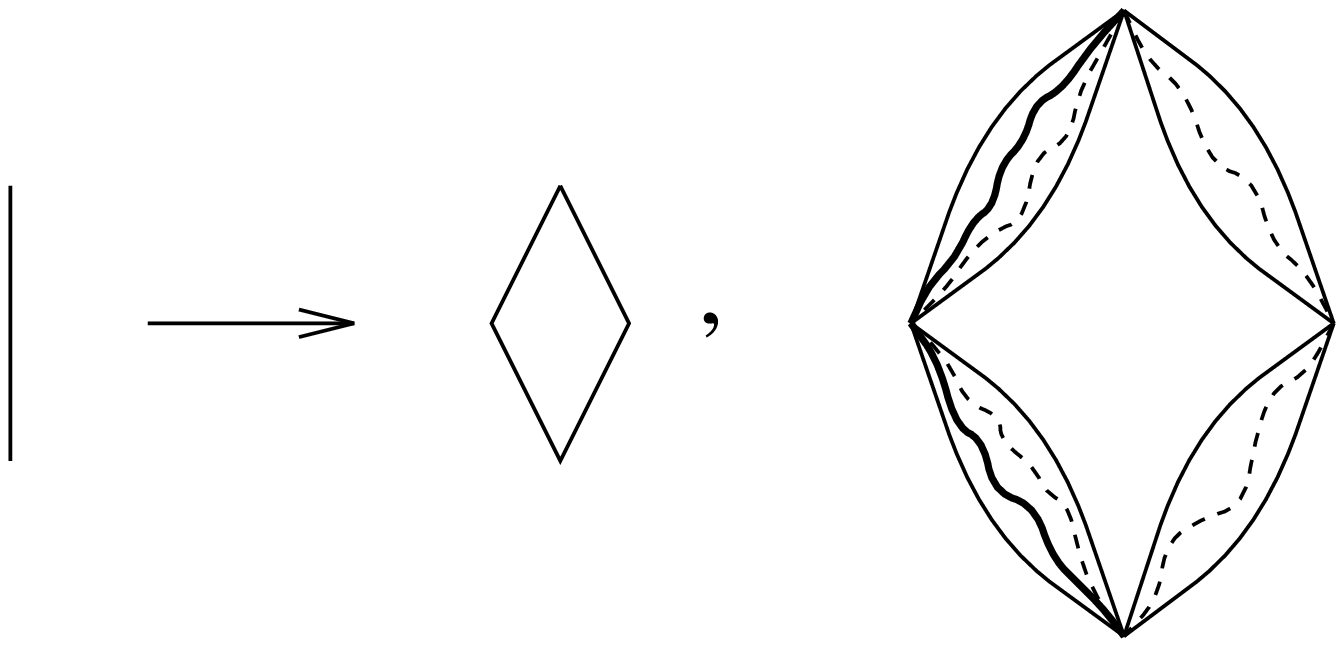}
\noindent{FIG. 3.} Construction rule of the DHL (level $n=0$ to level $n=1$)
and schematic picture of the lattice at level $n$, with the four
$n-1$ level units. A wall configuration (heavy) crossing the left
units, and two polymer configurations (dotted) are reported. With such
wall configuration eq. (6) applies.
\bigskip

By reinterpreting $\omega$ and $k$ as monomer fugacity and Boltzmann
factor for contacts, respectively,  our model  describes polymer
adsorption \cite{privman88}. Apart from a change in the ensemble
(${\cal Z}=\sum_X {\cal Z}_X$), the transfer matrices are the same.
Criticality ($\cal Z$ dominated by infinite length polymer
configurations) implies $\lambda_{max}=1$. We find that
$\lambda_{max}(1/2,k,\zeta_s)=1$ for all $k<k_c(1/2,\zeta_S)$, where
$k_c$ was defined above. For $k>k_c$ criticality  occurs at
$\omega<1/2$, indicating that the polymer is adsorbed
\cite{privman88}. Indeed $\omega=1/2$ marks criticality  for the
polymer in the bulk. The dependence of $k_c$ on $\zeta_S$ shows that
adsorption becomes more difficult with increasing roughness. One can
also show that the fraction of monomers adsorbed on the wall should
have the same singular behavior as $\overline{P_0}$, when moving at
$\omega=1/2$. Thus, like interface depinning, polymer adsorption
undergoes a change from second to first order upon increasing
$\zeta_S$.

To gain more insight into this change, we stick to polymer language
and consider a simplified model of adsorption defined on diamond
hierarchical lattice (DHL) (Fig. 3). Self--avoiding paths on DHL have
often been used to mimic directed polymers in 2$D$ \cite{derrida}. A
wall joining the two ends is obtained as follows: at level $n=0$ of
DHL construction the wall always coincides with the unique existing
bond. For $n=n_{max}>0$, the wall is determined by backward
iteration. Starting from $n_{max}$, at each level, $n$, we choose
whether the wall passes through the left or right DHL units of level
$n-1$, with probabilities $1-\Delta$ and $\Delta$, respectively, and
so on. If, e.g., we put $\Delta=0$, the process is deterministic and
we create a single wall coinciding with the left border of the
lattice. For $\Delta=1/2$ we generate with equal probability all
possible walls through the lattice. Given a wall, we consider a
polymer, with partition ${\cal Z}_n$, joining the ends of the lattice
and laying, e.g., to the right of the wall. As in eq. (2), $\omega$
and $k$ are step and wall contact fugacities, respectively. For
$\Delta=0$ we deal with a polymer attracted ($k>1$) by the left DHL
border, which plays the role of a ``flat'', deterministic substrate.
When $\Delta$ rises, the ``roughness'' of the now random wall
increases. Transverse hills and valleys are felt more and more by the
polymer. For a given wall, one can compute the polymer partition
function iteratively, using ${\cal Z}_{b,n+1}=2{\cal Z}_{b,n}^2$ and
\begin{equation}
{\cal Z}_{n+1}={\cal Z}_{n,1}{\cal Z}_{n,2}+{\cal Z}_{b,n}^2
\eqnum{6}
\end{equation}
or
\begin{equation}
{\cal Z}_{n+1}={\cal Z}_{n,1}{\cal Z}_{n,2}
\eqnum{7}
\end{equation}
Eq.(6) or (7) is chosen, according to whether, at level $n+1$, the
$n$--th level diamonds, 1 and 2, crossed by the wall, are the left or
the right ones, respectively. ${\cal Z}_{b,n}$ is clearly the ``bulk''
partition function on DHL at $n$--th level, in absence of wall.
Initial conditions are ${\cal Z}_0=k\omega$ and ${\cal
Z}_{b,0}=\omega$. Considering first $\Delta=0$, eq.(6) induces a two
parameters RG mapping by putting ${\cal Z}_{b,1}=\omega'$ and ${\cal
Z}_{1}={\omega}'k'$. Clearly the transformation of ${\cal Z}_{b,n}$
implies that  the bulk criticality condition is $\omega=1/2$ as in
the Euclidean case. The value $k=1$ is marginally unstable for
$k\gtrsim1$ and separates adsorbed ($k>1$) from desorbed ($k<1$)
regimes. Thus, for $\Delta=0$ a positive attraction is always
sufficient to adsorb the polymer. Eq. (6) applies for all $n$ and the
problem has a nontrivial fixed point with ${\cal Z}_b^*=1/2$ and
${\cal Z}^*=1/2$. $P_0(\omega=1/2, k,\Delta=0)$ can be extrapolated
by iteration. Due to marginality, $P_0$ starts rising with zero
slope, but continuously, for $k=1$ (Fig. 4). For $\Delta>0$ the
${\cal Z}_n$'s become random variables and we must iterate their
probability distribution, ${\cal P}_n$. This cannot be done exactly.
However, starting from ${\cal P}_0=\delta({\cal Z}-k\omega)$, we
follow the distribution at level $n$ by iteratively sampling it. From
a large number ($\simeq 10^6$) of ${\cal Z}$ values distributed
according to ${\cal P}_{n-1}$, we generate a sample distributed
according to ${\cal P}_n$ by choosing many pairs of ${\cal Z}$ values
and obtaining from each pair a new value of ${\cal Z}$ according to
eq.(6) or (7), with probabilities $1-\Delta$ and $\Delta$,
respectively. The resulting ${\cal Z}$'s constitute a sample of
${\cal P}_n$. This procedure could be iterated for $n\leq 40$ with
extremely high accuracy.

\vskip8.5truecm
\includegraphics{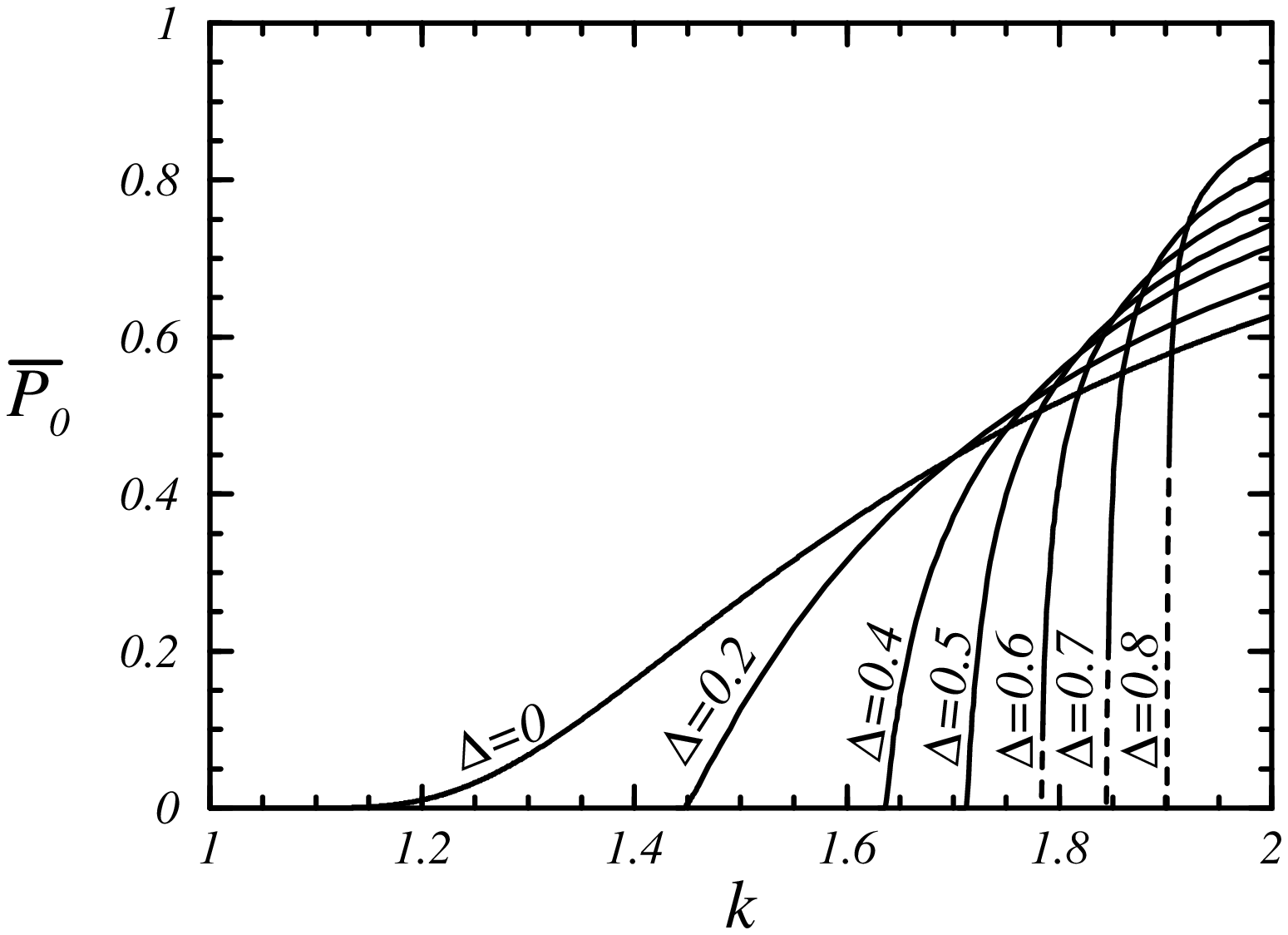}
\noindent{FIG. 4.} $\overline{P_0}$ on the DHL at $\omega=1/2$
for various $\Delta$ values. The dashed lines mark the discontinuities
for $\Delta>1/2$.
\bigskip

   $\overline{P_0}$ at $\omega=1/2$ is plotted as a function of $k$
in Fig. 4 for different $\Delta$'s. For $0<\Delta< 1/2$,
$\overline{P_0}$ rises continuously from zero as $c(k-k_c)$. Thus,
$\rho=1$ for $0<\Delta< 1/2$. The slope $c$ and $k_c$ are both
increasing with $\Delta$. Apart from $c=0$ at $\Delta=0$, due to the
accidental marginality, such behaviour reproduces what is observed in
2$D$ when $\zeta_S< 1/2$. For $\Delta<1/2$ the polymer is more
``rough'' than the wall. Only when $\Delta=1/2$ the latter has the
same freedom to develop through the DHL as a polymer has within the
``bulk''. On the other hand, $\Delta>1/2$ corresponds qualitatively
to $\zeta_S> 1/2$, because the polymer feels more and more the wall
limiting its options when developing through the DHL. For
$\Delta=1/2$ we have evidence that $c=\infty$, with a still
continuous transition. This infinite slope indeed anticipates a sharp
discontinuity in $\overline{P_0}$ for $\Delta>1/2$. So, the
hierarchical model contains ingredients reproducing, at least
qualitatively, the scenario emerging for the Euclidean model, and
gives a suggestive indication of the way in which continuous
transitions switch to first order at the expected threshold
$\zeta_S=1/2$.

    The dependence of $k_c$ on $\Delta$ mimics that on $\zeta_S$ in
2$D$, and further motivates the correspondence between $\Delta$ and
$\zeta_S$ in the two cases. We stress that, since all paths have the
same length, and there is no natural recipe for defining a transversal
distance on DHL, the notion of roughness must always be mediated in
some way: here we can link roughness to $\Delta$. Our hierarchical
model provides a remarkable example of the tricritical transition we
are dealing with in this paper, and allows an essentially exact
determination of its location and properties. Moreover, the threshold
we find at $\Delta=1/2$ is strongly suggestive of the precise
location of the tricritical point in the Euclidean case.

   Summarizing, in 2$D$, interface depinning or directed polymer
adsorption on rough substrate with $\zeta_S<1/2$ are continuous and
in the same universality as in the flat case. For $\zeta_S>1/2$
roughness is relevant and, moreover, the transitions acquire an
unusual, discontinuous nature. This result, not anticipated so far
\cite{li&kardar}, warns that some continuum approaches may not be
able to catch the correct physics of depinning from rough substrates.
We expect similar properties, and the same threshold $\zeta_S=1/2$,
for directed polymer adsorption on a self--affine surface in 3$D$.
Indeed, directed polymers have no upper critical dimension, and for
them $\zeta_0=1/2$ in all $d$. Although the single polymer adsorption
regime is not easily accessible experimentally, we believe that our
results should be relevant for stretched polymers \cite{ward}.

In 3$D$ the interface of a pure system typically has $\zeta_0=0$
\cite{lipowski}. We thus conjecture that depinning occurs
discontinuously as soon as $\zeta_S>0$, in cases when the interactions
are predominantly short--range, like in metallic systems. Our results
also bear on the observability of the recently predicted critical
wetting in the case of type--I superconductors \cite{okki}, which is
perhaps the most strict physical example of short--range
interface--substrate interactions.

Numerical calculations have been partly supported by CNR within the
CRAY project of Statistical Mechanics. We thank Joseph Indekeu for
valuable criticism and suggestions and Mehran Kardar for stimulating
discussions.

\end{document}